\documentstyle[12pt,aaspp,flushrt]{article}

\begin{document}

\title{Astrophysical Limits on the Evolution of Dimensionless \\
       Physical Constants over Cosmological Time}
\author{Lennox L. Cowie and Antoinette Songaila}
\affil{Institute for Astronomy, University of Hawaii, Honolulu, HI 96822}

\vskip 1in
\centerline{To be published in {\it Astrophysical Journal}, Nov.~10, 1995}

\begin{abstract}
We report new upper limits on any possible long-term time variation of the
ratio of the electron to proton masses, $(m_e/m_p)$, the fine-structure
constant, ($\alpha$), and the quantity $\alpha^2g_p(m_e/m_p)$, where $g_p$\
is the proton gyromagnetic ratio.  These limits are based on extremely high
precision observations of ${\rm H}_2$, ${\rm Si}^{3+}$, ${\rm C}^0$\ and
${\rm H}^0$\ in high-redshift quasar absorption lines.  They amount to 95\%
confidence ranges of $(-7.6\rightarrow 9.7) \times 10^{-14}~{\rm yr}^{-1}$\
for ($m_e/m_p$), $(-4.6\rightarrow 4.2)\times 10^{-14}~{\rm yr}^{-1}$\ for
$\alpha$\ and $(-2.2\rightarrow 4.2) \times 10^{-15}~{\rm yr}^{-1}$\ for
$\alpha^2g_p(m_e/m_p)$, where the elapsed time has been computed for a
cosmology with ${\rm H}_0 = 75~{\rm km\ s}^{-1}\ {\rm Mpc}^{-1}$\ and
${\rm q}_0 = 0.5$.
\end{abstract}


\newpage
\section{Introduction}

Current laboratory, geophysical and astrophysical constraints exclude any
significant evolution of most of the dimensionless physical constants over
cosmological time (Sisterna \& Vucetich 1990).  However, refining these
limits remains important as they provide constraints on theories such as
Kaluza-Klein and superstrings which allow solutions in which there can be
time variation of these constants.  We report improved astrophysical
constraints on the fine-structure constant, $\alpha$\ and the ratio of the
electron to proton masses, ($m_e/m_p$).

Savedoff (1956) first pointed out the possibility of using differential
measurements of redshifted lines in cosmologically distant objects to test
the evolution of certain of the dimensionless physical constants; since then
various measurements have been outlined and the observational constraints
greatly improved.  Recent summaries of these astrophysical measurements may
be found in Potekhin \& Varshalovich (1994) and Levshakov (1994).  In this
paper we describe the use of extremely high signal-to-noise and high
resolution ($R = 36,000$\ to 47,000) spectra of quasar absorption lines
obtained with the Keck 10~m telescope to tighten the evolutionary constraints
on $\alpha$, ($m_e/m_p$) and ($\alpha^2g_pm_e/m_p$) by roughly an order of
magnitude over previous measurements.  Note that, in translating limits to
time variation rates we have assumed a flat cosmology (${\rm q}_0 = 0.5$,
$\Lambda = 0$) with a Hubble constant of ${\rm H}_0 = 75~{\rm km\ s}^{-1}\
{\rm Mpc}^{-1}$.

\section{Constraints on ($m_e/m_p$)}

The presence of cosmological evolution in ($m_e/m_p$) can be tested by using
observations of molecular hydrogen in quasar absorption line systems,
assuming that all other dimensional constants, particularly the
fine-structure constant, do not vary in such a way as to compensate for the
variation in ($m_e/m_p$).  This was first pointed out by Thompson (1975), who
noted the different dependence of the electronic, vibrational and rotational
energy levels on ($m_e/m_p$).  There is, however, a general unavailability of
data at high redshift: to date, only one quasar, PKS~$0528-250$, is believed
to have molecular hydrogen in its spectrum, at a redshift of 2.811 (Chaffee,
Foltz \& Black 1988). Varshalovich and Levshakov (1993) have analyzed this
data to obtain a $1~\sigma$\ limit of $d(m_e/m_p)/dt \le 3 \times
10^{-13}~{\rm yr}^{-1}$, assuming a time difference of $1.3 \times
10^{10}~{\rm yr}$\ between $z = 2.811$\ and $z = 0$.

We have re-observed PKS~$0528-250$\ at much higher spectral resolution ($R =
36,000$) and signal-to-noise, obtaining a 5.4~hour spectrum with the Keck
Telescope which unambiguously confirms Chaffee et al.'s identification of
molecular hydrogen, as demonstrated in Fig.~1 which shows the 0--0 and
4--0 bands of the Lyman series.  An analysis of the physical properties of the
molecular hydrogen itself is given elsewhere (Songaila \& Cowie 1995).

Table~1 summarises the properties of those lines that are isolated
molecular hydrogen lines, based on their measured column densities and
FWHM.  In the Born-Oppenheimer approximation, the molecular hydrogen
energy levels can be written,
\begin{equation}
   E = E_{el} + \left({{m_e} \over {m_p}}\right)^{1/2}E_{vib} + \left({{m_e}
   \over {m_p}}\right)E_{rot}.
\end{equation}
So, the energy shift in any vibration-rotation transition $j$\ in the
Lyman series has the form,
\begin{equation}
   \Delta E_j = a_{el} + b_j \left({{m_e} \over {m_p}}\right)^{1/2} + c_j
   \left({{m_e} \over {m_p}}\right)
\end{equation}
and the difference in energy between two transitions is,
\begin{equation}
   (\Delta E_j - \Delta E_i) \sim b_{ji} \left({{m_e} \over {m_p}}\right)^{1/2}
   + c_{ji} \left({{m_e} \over {m_p}}\right).
\end{equation}
To lowest order, a change in $(m_e/m_p) \equiv \mu$\ induces a change in
$\Delta E_{i} - \Delta E_j$\ such that
\begin{equation}
   {{\delta\mu} \over {\mu}} \approx {{\delta v} \over {c}} \left({{2\Delta
   E_i} \over {\Delta E_i - \Delta E_j}}\right)
\end{equation}
where $\delta v$\ is the mean offset compared to the laboratory value, of the
energy difference between the two sets of lines, when that offset is
represented as a velocity difference.  While the above description is in fact
adequate for the analysis and gives a clear insight into the method, we have
in practice used the exact relationship,
\begin{equation}
   {{\delta v} \over {c}} = (k_i - k_j) \left({{\delta\mu} \over {\mu}}\right)
   = K_{ij} \left({{\delta\mu} \over {\mu}}\right)
\end{equation}
where the $k$\ coefficients have been tabulated by Varshalovich \& Levshakov
(1993).

A regression analysis based on the data of Table~1 then gives a best
fit value $\delta\mu /\mu = 8 \times 10^{-5}$\ and a 95\% confidence range of
$-5.5$\ to $7 \times 10^{-4}$.  The r.m.s. velocity spread is $2.5~{\rm km\
s}^{-1}$.  For our `standard' cosmology, the time to $z = 2.811$\ is $7.2
\times 10^9~{\rm yr}$\ and our 95\% confidence limit for $\dot{\mu}/\mu$\ is
$-7.6$\ to $9.7 \times 10^{-14}~{\rm yr}^{-1}$.  This is approximately an order
of magnitude improvement over the previous result.\goodbreak

\section{Constraints on $\alpha$}

The separation between the wavelength ($\lambda_1$) corresponding to the
transition $^2S_{1/2} \rightarrow ^2P_{3/2}$\ and that ($\lambda_2$)
corresponding to the transition $^2S_{1/2} \rightarrow ^2P_{1/2}$\ in an
alkaline ion is proportional to $\alpha^2$\ to lowest order in
$\alpha$\ (Bethe \& Salpeter 1977). Writing $\bar{\lambda} = (2/3)\lambda_1
+ \lambda_2$,
\begin{equation}
  {{(\lambda_1 - \lambda_2)} \over {\bar\lambda}} \equiv {{\Delta\lambda} \over
  {\bar\lambda}} \sim \alpha^2.
\end{equation}
So any change in $\alpha$\ will result in a corresponding change in
$\Delta\lambda$\ in the separation of the doublets in a high-$z$\ quasar (as
was first used by Bahcall \& Salpeter [1965] and Wolfe, Brown \& Roberts
[1976])
\begin{equation}
  {{\delta\alpha} \over {\alpha}} = {{\delta (\Delta\lambda)} \over
  {2\Delta\lambda}}
\end{equation}
\begin{equation}
  \left({{\delta\alpha} \over{\alpha}}\right) \approx \left({{\bar\lambda}
\over
  {2\Delta\lambda}}\right) \, \left({{\delta v} \over {c}}\right).
\end{equation}
The most recent analyses using this method (Potekhin \& Varshalovich 1994)
have used very large inhomogeneous samples of C~IV, N~V, O~VI, Mg~II, Al~III
and Si~IV doublets and obtained a 95\% confidence limit of $|
\alpha^{-1}(d\alpha /dz)| < 5.6 \times 10^{-4}$.

We have improved this analysis using a homogeneous sample of doublets
observed at high spectral reolution.  However, before proceeding further, it
is important to note that, for a given accuracy in determining $\Delta v/c$,
the sensitivity of determining $\Delta\alpha$\ is inversely proportional to
the relative splitting of the doublet $(\Delta\lambda /\bar\lambda)$.  For
this reason, Si~IV, the most widely spaced of the doublets ($\lambda\lambda
1393.755,\ 1402.770~{\rm\AA}$; Toresson 1960) is by far the most sensitive
probe of $\alpha$, followed by Al~III and Mg~II.  We have therefore
restricted our analysis to the Si~IV doublet.  Choosing a doublet with large
$(\Delta\lambda )/\bar\lambda$\ also alleviates a second problem, namely
that, for many of these doublets, the laboratory wavelength separation is not
known to very high precision.  This enters directly in the $(\delta
v/c)$\ term as a systematic uncertainty which is minimised for large
$(\Delta\lambda)/\bar\lambda$.  For Si~IV, Martin and Zalubas (1983) estimate
the uncertainty in Toresson's (1960) wavelengths at a maximum of $5~m\rm\AA$,
which, if adopted as an extreme upper bound to the uncertainty in the doublet
spacing, translates into an uncertainty of $1.0~{\rm km\ s}^{-1}$\ in $\Delta
v$, and currently imposes a fundamental limit of $2.7 \times 10^{-4}$\ on the
accuracy to which $\delta\alpha /\alpha$\ can be determined.

We have identified a number of Si~IV doublets in high-$z$\ quasar
absorption-line systems observed with the Keck HIRES spectrograph at $R =
36,000$.  The redshift systems used in the analysis were seen not only in
Si~IV but also in lines of other species, such as C~IV and Ly~$\alpha$;
therefore, the identifications do not depend solely on the spacing of the
Si~IV lines.  Table~2 lists the relative velocity shifts for Si~IV.  As is
illustrated in Fig.~2, the offsets have been measured by cross-correlating
the two members of the doublet.  The r.m.s. uncertainty in this procedure is
$1.1~{\rm km\ s}^{-1}$, determined from measurements of the more common C~IV
doublet.  The 95\% confidence limits inferred from Table~2 are $(\delta\alpha
/\alpha ) = (-2.2,1.6) \times 10^{-4}$, comparable to the systematic error
from the uncertainty in the laboratory wavelengths.  Combining the two errors
in quadrature, we have $|\delta\alpha/\alpha | < 3.5 \times 10^{-4}$\ and our
results imply $|\alpha^{-1}d\alpha/dz | < 1.1 \times 10^{-4}$, a factor of 5
lower than Potekhin and Varshalovich's (1994) result.  Further improvements
will require either new laboratory determinations of the Si~IV doublet
separation, or local observations of the Si~IV doublet spacing in the nearby
interstellar and intergalactic gas.  The later can be carried out by the
spectrographs on the Hubble Space Telescope.  To be useful for this purpose,
high resolution measurements with the Goddard High Resolution Spectrograph
would be required.  Current published values are not adequate.

\section{Constraints on $\alpha^2g_p(m_e/m_p)$}

The ratio of the frequencies of the hyperfine 21~cm absorption transition of
neutral hydrogen ($\nu_a$) to an optical resonance transition ($\nu_b$) has
the dependence
\begin{equation}
   {{\nu_a} \over {\nu_b}} \sim \alpha^2g_p\left({{m_e} \over {m_p}}\right)
   \equiv X
\end{equation}
so that evolution of this quantity will result in a difference in the measured
redshifts of 21~cm and optical absorption
\begin{equation}
   \delta z = z_{\rm opt} -z_{21} = (1+z)\left({{\delta x} \over {x}}\right).
\end{equation}
The current best limits on this quantity are given by Tubbs and Wolfe (1980),
who found $(1/x)(dx/dz) \le 1.1 \times 10^{-4}$\ for absorption in the quasar
Q1331+170.  The redshift of the hyperfine absorption in Q1331+170 is known to
very high precision ($z_{21} = 1.77642 \pm 2 \times 10^{-5}$; Wolfe \& Davis
1979) and the uncertainty in the Tubbs and Wolfe (1980) estimate arises
primarily in the imprecision of the optical redshift.  Recently, Songaila et
al.\ (1994) have presented high S/N observations of ${\rm C}^0$\ absorption
and fine-structure ${\rm C}^0$, finding $z_{\rm opt} = 1.77644 \pm 2 \times
10^{-5}$.  Because the ${\rm C}^0$\ arises in the same cloud components
responsible for the 21~cm absorption, the comparison of $z_{21}$\ and $z_{\rm
opt}$\ should be relatively secure against different kinematic structures
being present in the two measurements.  Combining the two redshifts, we find
\begin{equation}
   {{\delta X} \over {X}} = 7 \times 10^{-6} \pm 1.1 \times 10^{-5}
\end{equation}
corresponding to a 95\% confidence range for $(1/x)(dx/dz)$\ of $(-2.2,4.2)
\times 10^{-15}~{\rm yr}^{-1}$\ for our standard cosmology.

\section{Conclusion}

The three constraints presented here can be compared with existing laboratory
and astrophysical constraints.  The limits on $\alpha^2g_p(m_e/m_p)$\ currently
constitute the tightest astrophysical constraint on the evolution of these
dimensionless parameters, implying $|\dot{\alpha}/\alpha | \le 2.1 \times
10^{-15}~{\rm yr}^{-1}$\ and $|\dot{(m_e/m_p)}/(m_e/m_p)| \le 4.2 \times
10^{-15}~{\rm yr}^{-1}$\ at the 95\% confidence level, in the absence of
relative cancellation.  For $\dot{\alpha}/\alpha$, the astrophysical
measurement
is now comparable to the best local tests (Sisterna \& Vucetich 1990), which
give $|\dot{\alpha}/\alpha | \le 1.3 \times 10^{-15}~{\rm yr}^{-1}$\ (95\%
confidence), but, because of the larger timeline, may provide more stringent
constraints on models with non-linear evolution of $\alpha$.

\acknowledgements
We would like to thank Art Wolfe for many very helpful discussions.  The
authors were visiting astronomers at the  W. M. Keck Observatory, which is
jointly operated by the California Institute of Technology and the University
of California.

%
%  CONTAINS:   2 tables
%
\def\ang{~{\rm \AA}}
\def\kms{~{\rm km\ s}^{-1}}
\def\cm2{~{\rm cm}^{-2}}
\makeatletter
\def\jnl@aj{AJ}
\ifx\revtex@jnl\jnl@aj\let\tablebreak=\nl\fi
\makeatother
\newpage
%  TABLE 1
\begin{planotable}{crrr}
\tablewidth{300pt}
\tablecaption{Molecular Hydrogen Lines in PKS $0528-250$}
\tablehead{
\colhead{Line} & \colhead{$\lambda_{\rm vac}$\tablenotemark{a}} &
\colhead{$k$\tablenotemark{b}} & \colhead{$(\delta v/c)\
(\times 10^{-5})$\tablenotemark{c}~~~~~~}
}
\startdata
0--0 R(0) & 1108.128 & 7.79~~~~~ & 0.1~~~~~~ \nl
0--0 R(1) & 1108.633 & 8.25~~~~~ & $-$0.9~~~~~~ \nl
1--0 R(0) & 1092.195 & 0.72~~~~~ & $-$0.6~~~~~~ \nl
1--0 R(1) & 1092.732 & 1.22~~~~~ & 0.2~~~~~~ \nl
1--0 P91) & 1094.052 & 2.38~~~~~ & 0.7~~~~~~ \nl
1--0 R(2) & 1094.225 & 2.62~~~~~ & 1.2~~~~~~ \nl
1--0 P(2) & 1096.438 & 4.54~~~~~ & 1.3~~~~~~ \nl
2--0 R(1) & 1077.699 & $-$5.24~~~~~ & $-$1.1~~~~~~ \nl
2--0 P(1) & 1078.925 & $-$4.18~~~~~ & 0.3~~~~~~ \nl
2--0 R(2) & 1079.226 & $-$3.81~~~~~ & 0.2~~~~~~ \nl
2--0 P(2) & 1081.226 & $-$2.05~~~~~ & $-$0.9~~~~~~ \nl
3--0 P(1) & 1064.605 & $-$10.21~~~~~ & 0.5~~~~~~ \nl
3--0 R(2) & 1064.995 & $-$9.73~~~~~ & $-$0.4~~~~~~ \nl
3--0 P(2) & 1066.900 & $-$8.10~~~~~ & $-$1.2~~~~~~ \nl
4--0 R(0) & 1049.367 & $-$17.27~~~~~ & 0.8~~~~~~ \nl
4--0 R(1) & 1049.959 & $-$16.67~~~~~ & $-$1.0~~~~~~ \nl
4--0 P(1) & 1051.032 & $-$15.76~~~~~ & $-$0.5~~~~~~ \nl
4--0 R(2) & 1051.498 & $-$15.18~~~~~ & 1.4~~~~~~ \nl
4--0 P(2) & 1053.284 & $-$13.67~~~~~ & 0.0~~~~~~ \nl
\tablenotetext{a}{Dabrowski 1984}
\tablenotetext{b}{Potekhin \& Varshalovich 1994}
\tablenotetext{c} {Frame of reference is such that $\langle
\delta v/c\rangle = 0$}
\end{planotable}
\clearpage

\newpage
%  TABLE 2
\begin{planotable}{crr}
\tablewidth{250pt}
\tablecaption{Si~IV Doublet Separations}
\tablehead{
\colhead{Quasar} & \colhead{$z$} & \colhead{$\delta v$}
}
\startdata
0302$-$003 & 2.785 & 0.8~~~ \nl
0528$-$250 & 2.813 & 0.5~~~ \nl
0528$-$250 & 2.810 & 0.4~~~ \nl
0528$-$250 & 2.672 & $-$2.1~~~ \nl
1206+119 & 3.021 & $-$0.5~~~ \nl
2000$-$330 & 3.551 & $-$1.5~~~ \nl
2000$-$330 & 3.548 & 1.1~~~ \nl
2000$-$330 & 3.332 & 2.3~~~ \nl
2000$-$330 & 3.191 & $-$2.2~~~ \nl
\end{planotable}
\clearpage

\begin{figure}
\caption{Spectra of the 0--0 and 4--0 vibration transitions of the Lyman series
of molecular hydrogen at $z = 2.811$\ in the quasar PKS~$0528-250$.  The
spectrum is shown in the rest frame with the dashed lines indicating the
wavelengths of the $J \le 2$\ transitions in each band.  The remaining
(generally broader) lines in the spectra are primarily Lyman alpha forest lines
of intergalactic neutral hydrogen.}
\end{figure}

\begin{figure}
\caption{Cross correlation of the two lines of the Si~IV doublet for a cloud
in the spectrum of the quasar Q$0302-003$.  The quantity $(1-S(\lambda))$ is
plotted versus $v$, where $S$\ is the normalised spectrum for the stronger
line, and the quantity $2(1-S(\lambda))$\ (dashed line) is overplotted versus
$v - \Delta v$\ for the weaker line.  Here $\Delta v$\ is the measured
offset.  The vertical dashed lines show the velocity range over which the
cross correlation function was computed.}
\end{figure}

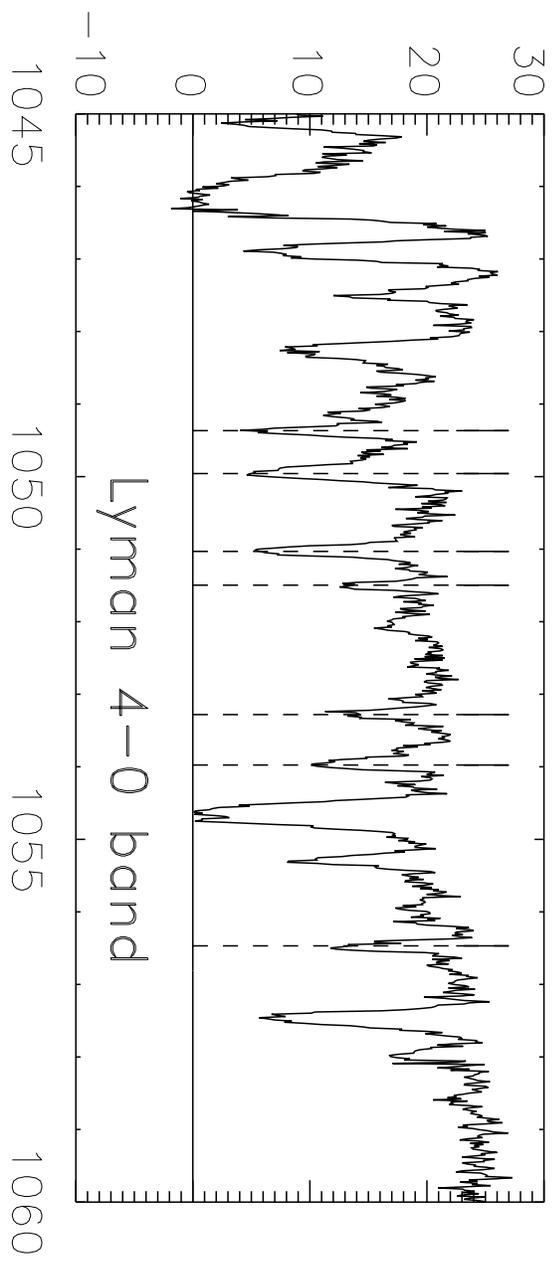
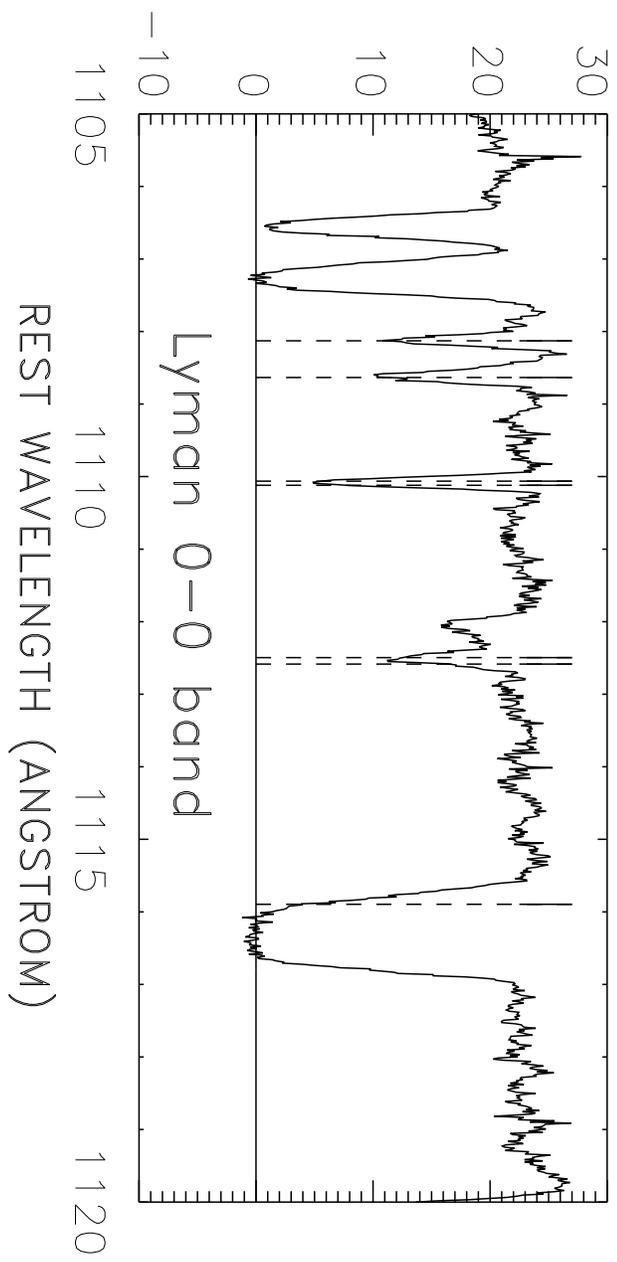
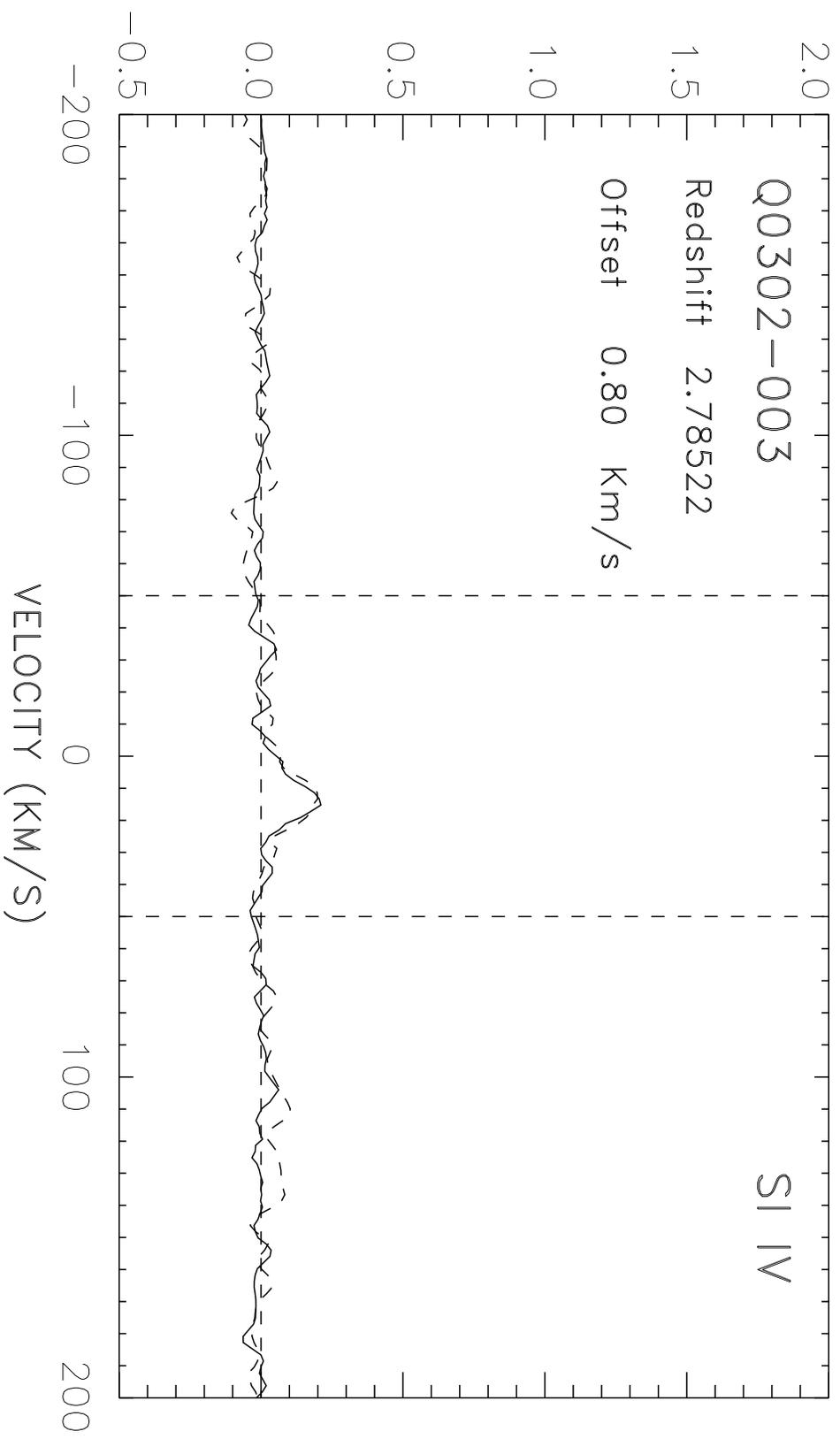
\end{document}